\documentclass[12pt]{article}

\makeatletter

\@addtoreset{equation}{section}
\makeatother

\newtheorem{definition}{Definition}[section]
\newtheorem{theorem}{Theorem}[section]

\newtheorem{lemma}{Lemma}[section]
\newtheorem{remark}{Remark}[section]

\usepackage{amssymb}
\usepackage{latexsym}

\begin{document} 

\title{Uncertainty Relation on Wigner-Yanase-Dyson skew information, II}
\author{Kenjiro Yanagi\thanks{
Division of Applied Mathematical Science,
Graduate School of Science and Engineering,
Yamaguchi University,
Ube, 755-8611 Japan. 
E-mail: {\tt yanagi@\allowbreak
yamaguchi-u.\allowbreak
ac.\allowbreak
jp},  
This research was partially supported by the Ministry of Education, Science, Sports and Culture, Grant-in-Aid for 
Scientific Research (C), 20540175}}
\date{}

\maketitle

\begin{flushleft}
{\bf Abstract.} We give a trace inequality related to the uncertainty relation 
of generalized Wigner-Yanase-Dyson skew information which includes our result in \cite{Ya:WYD}. 
\end{flushleft}

\begin{flushleft}
\end{flushleft}

\section{Introduction}

Wigner-Yanase skew information
\begin{eqnarray*}
I_{\rho}(H) & = & \frac{1}{2}Tr \left[ \left( i \left[ \rho^{1/2},H \right] \right)^2 \right] \\
& = & Tr[\rho H^2]-Tr[\rho^{1/2}H\rho^{1/2}H] 
\end{eqnarray*}
was defined in \cite{WY:inf}. This quantity can be considered as a kind of 
the degree for non-commutativity between a quantum state $\rho$ and an 
observable $H$. Here we denote the commutator by $[X,Y] = XY-YX$. This 
quantity was generalized by Dyson 
\begin{eqnarray*}
I_{\rho,\alpha}(H) & = & \frac{1}{2}Tr[(i[\rho^{\alpha},H])(i[\rho^{1-\alpha},H])] \\
& = & Tr[\rho H^2]-Tr[\rho^{\alpha}H\rho^{1-\alpha}H],  \alpha \in [0,1]
\end{eqnarray*}
which is known as the Wigner-Yanase-Dyson skew information. It is famous that 
the convexity of $I_{\rho,\alpha}(H)$ with respect to $\rho$ was successfully 
proven by E.H.Lieb in \cite{Li:convex}. 
And also this quantity was generalized by Cai and Luo
\begin{eqnarray*}
&   & I_{\rho,\alpha,\beta}(H) \\
& = & \frac{1}{2}Tr[(i[\rho^{\alpha},H])(i[\rho^{\beta},H])\rho^{1-\alpha-\beta}] \\
& = & \frac{1}{2} \{ Tr[\rho H^2]+Tr[\rho^{\alpha+\beta}H\rho^{1-\alpha-\beta}H]-Tr[\rho^{\alpha}H\rho^{1-\alpha}H]-Tr[\rho^{\beta}H\rho^{1-\beta}H] \},
\end{eqnarray*}
where $\alpha, \beta \geq 0, \alpha+\beta \leq 1$. The convexity of $I_{\rho,\alpha,\beta}(H)$ with respect to $\rho$ 
was proven by Cai and Luo in \cite{CaLu:convex} under some restrictive condition. 
From the physical point of view, 
an observable $H$ is generally considered to be an unbounded opetrator, 
however in the present paper, unless otherwise stated, we consider 
$H \in B({\cal H})$ represents the set of all bounded linear operators on 
the Hilbert space ${\cal H}$, as a mathematical interest. We also denote 
the set of all self-adjoint operators (observables) by ${\cal L}_h({\cal H})$ 
and the set of all density operators (quantum states) by ${\cal S}({\cal H})$ 
on the Hilbert space ${\cal H}$. The relation between the Wigner-Yanase skew 
information and the uncertainty relation was studied in \cite{LuZh:skew}. 
Moreover the relation between the Wigner-Yanase-Dyson skew information and the 
uncertainty relation was studied in \cite{Ko:matrix, YaFuKu:gen}. In our paper 
\cite{YaFuKu:gen} and \cite{Ya:WYD}, we defined a generalized skew information and then derived 
a kind of an uncertainty relations. In the section 2, we discuss various 
properties of Wigner-Yanase-Dyson skew information. 
In section 3, we give an uncertainty relation of generalized Wigner-Yanase-Dyson skew information.

\section{Trace inequality of Wigner-Yanase-Dyson skew information}

We review the relation between the Wigner-Yanase skew information and the uncertainty relation. 
In quantum mechanical system, the expectation value of an observable $H$ 
in a quantum state $\rho$ is expressed by $Tr[\rho H]$. It is natural 
that the variance for a quantum state $\rho$ and an observable $H$ is defined by 
$V_{\rho}(H) = Tr[\rho (H-Tr[\rho H]I)^2] = Tr[\rho H^2]-Tr[\rho H]^2$. 
It is famous that we have 
\begin{equation}
V_{\rho}(A)V_{\rho}(B) \geq \frac{1}{4}|Tr[\rho [A,B]]|^2 
\label{eq:num2-1}
\end{equation}
for a quantum state $\rho$ and two observables $A$ and $B$. The further strong results 
was given by Schrodinger 
$$
V_{\rho}(A)V_{\rho}(B)-|Cov_{\rho}(A,B)|^2 \geq \frac{1}{4}|Tr[\rho [A,B]]|^2, 
$$
where the covariance is defined by 
$Cov_{\rho}(A,B) = Tr[\rho (A-Tr[\rho A]I)(B-Tr[\rho B]I)]$. 
However, the uncertainty relation for the Wigner-Yanase skew information failed. 
(See \cite{LuZh:skew, Ko:matrix, YaFuKu:gen}) 
$$
I_{\rho}(A) I_{\rho}(B) \geq \frac{1}{4}|Tr[\rho [A,B]]|^2.
$$
Recently, S.Luo introduced the quantity $U_{\rho}(H)$ representing a quantum uncertainty 
excluding the classical mixture: 
\begin{equation}
U_{\rho}(H) = \sqrt{V_{\rho}(H)^2-(V_{\rho}(H)-I_{\rho}(H))^2}, 
\label{eq:num2-2}
\end{equation}
then he derived the uncertainty relation on $U_{\rho}(H)$ in \cite{Lu:Hei}: 
\begin{equation}
U_{\rho}(A) U_{\rho}(B) \geq \frac{1}{4}|Tr[\rho [A,B]]|^2.
\label{eq:num2-3}
\end{equation}
Note that we have the following relation 
\begin{equation}
0 \leq I_{\rho}(H) \leq U_{\rho}(H) \leq V_{\rho}(H). 
\label{eq:num2-4}
\end{equation}
The inequality (\ref{eq:num2-3}) is a refinement of the inequality (\ref{eq:num2-1}) 
in the sense of (\ref{eq:num2-4}). 
In \cite{Ya:WYD}, we studied one-parameter extended inequality for the inequality (\ref{eq:num2-3}). 

\begin{definition}
For $0 \leq \alpha \leq 1$, a quantum state $\rho$ and an observable $H$, we define 
the Wigner-Yanase-Dyson skew information 
\begin{eqnarray}
I_{\rho,\alpha}(H) & = & \frac{1}{2}Tr[(i [\rho^{\alpha},H_0])(i [\rho^{1-\alpha},H_0])] \nonumber \\
& = & Tr[\rho H_0^2]-Tr[\rho^{\alpha}H_0\rho^{1-\alpha}H_0] 
\label{eq:num2-5}
\end{eqnarray}
and we also define 
\begin{eqnarray}
J_{\rho,\alpha}(H) & = & \frac{1}{2}Tr[\{ \rho^{\alpha},H_0 \} \{ \rho^{1-\alpha},H_0 \}]  \nonumber \\
& = & Tr[\rho H_0^2]+Tr[\rho^{\alpha}H_0 \rho^{1-\alpha}H_0], 
\label{eq:num2-6}
\end{eqnarray}
where $H_0 = H-Tr[\rho H]I$ and we denote the anti-commutator by $\{ X,Y \} = XY+YX$.  
\label{def:definition2-1}
\end{definition}

Note that we have 
$$
\frac{1}{2}Tr[(i [\rho^{\alpha},H_0])(i [\rho^{1-\alpha},H_0])] = \frac{1}{2}Tr[(i [\rho^{\alpha},H])(i [\rho^{1-\alpha},H])]
$$
but we have 
$$
\frac{1}{2}Tr[\{ \rho^{\alpha},H_0 \} \{ \rho^{1-\alpha},H_0 \}] \neq \frac{1}{2}Tr[\{\rho^{\alpha},H \} \{ \rho^{1-\alpha},H \}]. 
$$
Then we have the following inequalities: 
\begin{equation}
I_{\rho,\alpha}(H) \leq I_{\rho}(H) \leq J_{\rho}(H) \leq J_{\rho,\alpha}(H), 
\label{eq:num2-7}
\end{equation}
since we have $Tr[\rho^{1/2}H\rho^{1/2}H] \leq Tr[\rho^{\alpha}H\rho^{1-\alpha}H]$. 
(See \cite{Bo:some,Fu:trace} for example.) If we define 
\begin{equation}
U_{\rho,\alpha}(H) = \sqrt{V_{\rho}(H)^2-(V_{\rho}(H)-I_{\rho,\alpha}(H))^2}, 
\label{eq:num2-8}
\end{equation}
as a direct generalization of Eq.(\ref{eq:num2-2}), then we have 
\begin{equation}
0 \leq I_{\rho,\alpha}(H) \leq U_{\rho,\alpha}(H) \leq U_{\rho}(H) 
\label{eq:num2-9}
\end{equation}
due to the first inequality of (\ref{eq:num2-7}). We also have 
$$
U_{\rho,\alpha}(H) = \sqrt{I_{\rho,\alpha}(H) J_{\rho,\alpha}(H)}. 
$$
From the inequalities (\ref{eq:num2-4}),(\ref{eq:num2-8}),(\ref{eq:num2-9}), our situation is that we have 
$$
0 \leq I_{\rho,\alpha}(H) \leq I_{\rho}(H) \leq U_{\rho}(H)
$$
and
$$
0 \leq I_{\rho,\alpha}(H) \leq U_{\rho,\alpha}(H) \leq U_{\rho}(H).
$$
We gave the following uncertainty relation with respect to $U_{\rho,\alpha}(H)$ as a direct 
generalization of the inequality (\ref{eq:num2-3}). 

\begin{theorem}[\cite{Ya:WYD}]
For $0 \leq \alpha \leq 1$, a quantum state $\rho$ and observablea $A,B$, 
\begin{equation}
U_{\rho,\alpha}(A) U_{\rho,\alpha}(B) \geq \alpha(1-\alpha)|Tr[\rho [A,B]]|^2.
\label{eq:num2-10}
\end{equation}
\label{th:theorem2-1}
\end{theorem}

Now we define the two parameter extensions of Wigner-Yanase skew information and 
give an uncertainty relation under some conditions in the next section. 

\begin{definition}
For $\alpha, \beta \geq 0$, a quantum state $\rho$ and an observable $H$, 
we define the generalized Wigner-Yanase-Dyson skew information
\begin{eqnarray*}
&   & I_{\rho,\alpha,\beta}(H) \\
& = & \frac{1}{2}Tr\left[(i[\rho^{\alpha},H_0])(i[\rho^{\beta},H_0])\rho^{1-\alpha-\beta} \right] \\
& = & \frac{1}{2}\{ Tr[\rho H_0^2]+Tr[\rho^{\alpha+\beta}H_0 \rho^{1-\alpha-\beta}H_0]-Tr[\rho^{\alpha}H_0 \rho^{1-\alpha}H_0]-Tr[\rho^{\beta}H_0 \rho^{1-\beta}H_0] \} 
\end{eqnarray*}
and we define 
\begin{eqnarray*}
&   & J_{\rho,\alpha,\beta}(H) \\
& = & \frac{1}{2}Tr \left[(i \{\rho^{\alpha},H_0 \})(i \{\rho^{\beta},H_0 \}) \rho^{1-\alpha-\beta} \right] \\
& = & \frac{1}{2}\{ Tr[\rho H_0^2]+Tr[\rho^{\alpha+\beta}H_0 \rho^{1-\alpha-\beta}H_0]+Tr[\rho^{\alpha}H_0 \rho^{1-\alpha}H_0]+Tr[\rho^{\beta}H_0 \rho^{1-\beta}H_0] \},
\end{eqnarray*}
where $H_0 = H-Tr[\rho H]I$ and we denote the anti-commutator by $\{X,Y \} = XY+YX$. We remark that $\alpha+\beta = 1$ implies 
$I_{\rho,\alpha}(H) = I_{\rho,\alpha,1-\alpha}(H)$ and $J_{\rho,\alpha}(H) = J_{\rho,\alpha,1-\alpha}(H)$.
We also define 
$$
U_{\rho,\alpha,\beta}(H) = \sqrt{I_{\rho,\alpha,\beta}(H) J_{\rho,\alpha,\beta}(H)}.
$$
\label{def:definition2-2}
\end{definition}

\section{Main Theorem}

In this section we assume that $\rho$ is invertible density matrix and $A, B$ are Hermitian matrices. 
We also assume that $\alpha, \beta \geq 0$ do not necessarily satisfy the condition $\alpha+ \beta \leq 1$. 
We give the main theorem as follows; 

\begin{theorem}
For $\alpha, \beta \geq 0$ and $\alpha + \beta \geq 1$ or $\alpha + \beta \leq \frac{1}{2}$, 
\begin{equation}
U_{\rho,\alpha, \beta}(A)U_{\rho,\alpha,\beta}(B) \geq \alpha \beta|Tr[\rho [A,B]]|^2.
\label{eq:num3-1}
\end{equation}
\label{th:theorem3-1}
\end{theorem}

\vspace{0.8cm}
\noindent
We use the several lemmas to prove the theorem \ref{th:theorem3-1}. By spectral decomposition, there exists an orthonormal basis 
$\{ \phi_1, \phi_2,\ldots, \phi_n \}$ consisting of eigenvectors of $\rho$. Let $\lambda_1, \lambda_2,\ldots, \lambda_n$ 
be the corresponding eigenvalues, where $\sum_{i=1}^n \lambda_i = 1$ and $\lambda_i > 0$. 
Thus, $\rho$ has a spectral representation 
\begin{equation}
\rho = \sum_{i=1}^n \lambda_i |\phi_i \rangle \langle \phi_i|.
\label{eq:num3-2}
\end{equation}

We use the notation $f_{\alpha}(x,y) = x^{\alpha}y^{1-\alpha}+x^{1-\alpha}y^{\alpha}$. Then we have the following lemmas.

\begin{lemma}
\begin{eqnarray*}
I_{\rho,\alpha,\beta}(H) & = & \frac{1}{2}\sum_{i<j} \{ \lambda_i+\lambda_j+f_{\alpha+\beta}(\lambda_i,\lambda_j)-f_{\alpha}(\lambda_i,\lambda_j)-f_{\beta}(\lambda_i,\lambda_j) \} |\langle \phi_i|H_0|\phi_j \rangle|^2.
\end{eqnarray*}
\label{lem:lemma3-1}
\end{lemma}

\begin{flushleft}
{\bf Proof of Lemma \ref{lem:lemma3-1}.} By (\ref{eq:num3-2}), 
\end{flushleft}
$$
\rho H_0^2 = \sum_{i=1}^n \lambda_i|\phi_i\rangle \langle \phi_i|H_0^2.
$$
Then 
\begin{equation}
Tr[\rho H_0^2]  =  \sum_{i=1}^n \lambda_i \langle \phi_i|H_0^2|\phi_i \rangle = \sum_{i=1}^n \lambda_i \| H_0|\phi_i \rangle \|^2. 
\label{eq:num3-3}
\end{equation}
Since 
$$
\rho^{\alpha}H_0 = \sum_{i=1}^n \lambda_i^{\alpha} |\phi_i \rangle \langle \phi_i|H_0
$$
and 
$$
\rho^{1-\alpha}H_0 = \sum_{i=1}^n \lambda_i^{1-\alpha} |\phi_i \rangle \langle \phi_i|H_0,
$$
we have 
$$
\rho^{\alpha}H_0 \rho^{1-\alpha}H_0 = \sum_{i,j=1}^n \lambda_i^{\alpha}\lambda_j^{1-\alpha}|\phi_i \rangle  
\langle \phi_i|H_0|\phi_j \rangle \langle \phi_j|H_0. 
$$
Thus 
\begin{eqnarray}
Tr[\rho^{\alpha}H_0 \rho^{1-\alpha}H_0] & = & \sum_{i,j=1}^n \lambda_i^{\alpha}\lambda_j^{1-\alpha} \langle \phi_i|H_0|\phi_j \rangle \langle \phi_j|H_0|\phi_i \rangle \nonumber \\
& = & \sum_{i,j=1}^n \lambda_i^{\alpha}\lambda_j^{1-\alpha} | \langle \phi_i|H_0|\phi_j \rangle |^2. \label{eq:num3-4}
\end{eqnarray}
By the similar calculations we have 
\begin{eqnarray}
Tr[\rho^{\beta}H_0 \rho^{1-\beta}H_0] & = & \sum_{i,j=1}^n \lambda_i^{\beta}\lambda_j^{1-\beta} \langle \phi_i|H_0|\phi_j \rangle \langle \phi_j|H_0|\phi_i \rangle \nonumber \\
& = & \sum_{i,j=1}^n \lambda_i^{\beta}\lambda_j^{1-\beta} | \langle \phi_i|H_0|\phi_j \rangle |^2. \label{eq:num3-5}
\end{eqnarray}
\begin{eqnarray}
Tr[\rho^{\alpha+\beta}H_0 \rho^{1-\alpha-\beta}H_0] & = & \sum_{i,j=1}^n \lambda_i^{\alpha+\beta}\lambda_j^{1-\alpha-\beta} \langle \phi_i|H_0|\phi_j \rangle \langle \phi_j|H_0|\phi_i \rangle \nonumber \\
& = & \sum_{i,j=1}^n \lambda_i^{\beta}\lambda_j^{1-\beta} | \langle \phi_i|H_0|\phi_j \rangle |^2. \label{eq:num3-6}
\end{eqnarray}
From (\ref{eq:num2-5}), (\ref{eq:num3-3}), (\ref{eq:num3-4}), (\ref{eq:num3-5}), (\ref{eq:num3-6}), 
\begin{eqnarray*}
&   & I_{\rho,\alpha,\beta}(H) \\
& = & \frac{1}{2} \sum_{i,j}(\lambda_i+\lambda_i^{\alpha+\beta}\lambda_j^{1-\alpha-\beta}-\lambda_i^{\alpha}\lambda_j^{1-\alpha}-\lambda_i^{\beta}\lambda_j^{1-\beta}) | \langle \phi_i|H_0|\phi_j \rangle |^2 \\
& = & \frac{1}{2} \sum_{i<j} (\lambda_i+\lambda_i-\lambda_i-\lambda_i) | \langle \phi_i|H_0|\phi_i \rangle |^2 \\
&   & +\frac{1}{2} \sum_{i<j} (\lambda_i+\lambda_i^{\alpha+\beta}\lambda_j^{1-\alpha-\beta}-\lambda_i^{\alpha}\lambda_j^{1-\alpha}-\lambda_i^{\beta}\lambda_j^{1-\beta}) | \rangle \phi_i|H_0| \phi_j \rangle |^2 \\
&   & +\frac{1}{2} \sum_{i<j} (\lambda_j+\lambda_j^{\alpha+\beta}\lambda_i^{1-\alpha-\beta}-\lambda_j^{\alpha}\lambda_i^{1-\alpha}-\lambda_j^{\beta}\lambda_i^{1-\beta}) | \rangle \phi_j|H_0| \phi_i \rangle |^2 \\
& = & \frac{1}{2} \sum_{i<j} (\lambda_i+\lambda_j+f_{\alpha+\beta}(\lambda_i,\lambda_j)-f_{\alpha}(\lambda_i,\lambda_j)-f_{\beta}(\lambda_i,\lambda_j)) | \langle \phi_i|H_0| \phi_j \rangle |^2. 
\end{eqnarray*}
\ \hfill $\Box$

\begin{lemma}
$$
J_{\rho,\alpha,\beta}(H) \geq \sum_{i<j} (\lambda_i+\lambda_j+f_{\alpha+\beta}(\lambda_i,\lambda_j)+f_{\alpha}(\lambda_i,\lambda_j)+f_{\beta}(\lambda_i,\lambda_j)) | \langle \phi_i|H_0|\phi_j \rangle |^2. 
$$
\label{lem:lemma3-2}
\end{lemma}

\begin{flushleft}
{\bf Proof of Lemma \ref{lem:lemma3-2}.}  By (\ref{eq:num2-6}), (\ref{eq:num3-3}), (\ref{eq:num3-4}), (\ref{eq:num3-5}), 
(\ref{eq:num3-6}), we have 
\end{flushleft}
\begin{eqnarray*}
&   & J_{\rho,\alpha}(H) \\
& = & \frac{1}{2} \sum_{i,j}(\lambda_i+\lambda_i^{\alpha+\beta}\lambda_j^{1-\alpha-\beta}+\lambda_i^{\alpha}\lambda_j^{1-\alpha}+\lambda_i^{\beta}\lambda_j^{1-\beta}) | \langle \phi_i|H_0| \phi_j \rangle |^2 \\
& = & \frac{1}{2} \sum_i (\lambda_i+\lambda_i+\lambda_i+\lambda_i) | \langle \phi_i|H_0| \phi_i \rangle |^2 \\
&   & + \frac{1}{2} \sum_{i<j}(\lambda_i+\lambda_i^{\alpha+\beta}\lambda_j^{1-\alpha-\beta}+\lambda_i^{\alpha}\lambda_j^{1-\alpha}+\lambda_i^{\beta}\lambda_j^{1-\beta}) | \langle \phi_i|H_0| \phi_j \rangle |^2 \\
&   & + \frac{1}{2} \sum_{i<j}(\lambda_j+\lambda_j^{\alpha+\beta}\lambda_i^{1-\alpha-\beta}+\lambda_j^{\alpha}\lambda_i^{1-\alpha}+\lambda_j^{\beta}\lambda_i^{1-\beta}) | \langle \phi_j|H_0| \phi_i \rangle |^2 \\
& = & 2 \sum_i \lambda_i | \langle \phi_i|H_0| \phi_i \rangle |^2 \\
&   & + \frac{1}{2} \sum_{i<j}(\lambda_i+\lambda_j+f_{\alpha+\beta}(\lambda_i,\lambda_j)+f_{\alpha}(\lambda_i,\lambda_j)+f_{\beta}(\lambda_i,\lambda_j) | \langle \phi_i|H_0| \phi_j \rangle |^2 \\
& \geq & \frac{1}{2} \sum_{i<j} ( \lambda_i+\lambda_j+f_{\alpha+\beta}(\lambda_i,\lambda_j)+f_{\alpha}(\lambda_i,\lambda_j)+f_{\beta}(\lambda_i,\lambda_j) | \langle \phi_i|H_0| \phi_j \rangle |^2. 
\end{eqnarray*}
\ \hfill $\Box$

\begin{lemma}
For any $t > 0$ and $\alpha, \beta \geq 0, \alpha + \beta \geq 1$ or $\alpha + \beta \leq \frac{1}{2}$, the following inequality holds; 
\begin{equation}
(t^{1-\alpha-\beta}+1)^2(t^{2\alpha}-1)(t^{2\beta}-1) \geq 16 \alpha \beta (t-1)^2.
\label{eq:num3-7} 
\end{equation}
\label{lem:lemma3-3}
\end{lemma}

\vspace{1cm}
\noindent
{\bf Proof of Lemma \ref{lem:lemma3-3}.}  It is sufficient to prove (\ref{eq:num3-7}) for $t \geq 1$ and $\alpha, \beta \geq 0, \alpha + \beta \geq 1$ or $\alpha + \beta \leq \frac{1}{2}$.  
By Lemma 3.3 in \cite{Ya:WYD} we have  for $0 \leq p \leq 1$ and $s \geq 1$,  
$$
(1-2p)^2(s-1)^2-(s^p-s^{1-p})^2 \geq 0.
$$
Then we can rewrite as follows; 
$$
(s^{2p}-1)(s^{2(1-p)}-1) \geq 4p(1-p)(s-1)^2.
$$
We assume that $\alpha, \beta \geq 0$.  We put $p =\alpha/(\alpha+\beta)$ and $s^{1/(\alpha+\beta)} = t$. Then 
$$
(t^{2\alpha}-1)(t^{2\beta}-1) \geq \frac{4\alpha \beta}{(\alpha+\beta)^2}(t^{\alpha+\beta}-1)^2.
$$
Then we have 
\begin{equation}
(t^{1-\alpha-\beta}+1)^2(t^{2\alpha}-1)(t^{2\beta}-1) \geq \frac{4\alpha \beta}{(\alpha+\beta)^2}(t^{1-\alpha-\beta}+1)^2(t^{\alpha+\beta}-1)^2.
\label{eq:num3-8}
\end{equation}
We put $\alpha+\beta = k$ and $f(t) = (t^{1-k}+1)(t^k-1)-2k(t-1)$. Then 
\begin{eqnarray*}
f^{'}(t) & = & (1-k)t^{-k}(t^k-1)+k(t^{1-k}+1)t^{k-1}-2k \\
         & = & (1-k)(1-t^{-k})+k(1+t^{k-1})-2k.
\end{eqnarray*}
and
\begin{eqnarray*}
f^{''}(t) & = & (1-k)kt^{-k-1}+k(k-1)t^{k-2} \\
          & = & k(k-1)(t^{k-2}-t^{-k-1}). 
\end{eqnarray*} 
When $k =\alpha+\beta \geq 1$ or $ k = \alpha+\beta \leq \frac{1}{2}$, it is easy to show that $f^{''}(t) \geq 0$ for $t \geq 1$. 
Since $f^{'}(1) = 0$, we have $f^{'}(t) \geq 0$ for $t \geq 1$.  And since $f(1) = 0$, we have 
$f(t) \geq 0$ for $t \geq 1$. 
Hence we have for $\alpha+\beta \geq 1$ or $\alpha+\beta \leq \frac{1}{2}$, 
$$
(t^{1-\alpha-\beta}+1)(t^{\alpha+\beta}-1) \geq 2(\alpha+\beta)(t-1).
$$
It follows from (\ref{eq:num3-8}) that we get 
$$
(t^{1-\alpha-\beta}+1)^2(t^{2\alpha}-1)(t^{2\beta}-1) \geq 16\alpha \beta (t-1)^2.
$$
\ \hfill $\Box$

\vspace{1cm}
\noindent
{\bf Proof of Theorem \ref{th:theorem3-1}.} Since 
\begin{eqnarray*}
&   & (t^{1-\alpha-\beta}+1)^2(t^{2\alpha}-1)(t^{2\beta}-1) \\
& = & (t+1+t^{\alpha+\beta}+t^{1-\alpha-\beta})^2-(t^{\alpha}+t^{1-\alpha}+t^{\beta}+t^{1-\beta})^2, 
\end{eqnarray*}
we put $\displaystyle{t = \frac{\lambda_i}{\lambda_j}}$ 
in (\ref{eq:num3-7}).  Then we have
$$
\left\{ \frac{\lambda_i}{\lambda_j}+1+\left(\frac{\lambda_i}{\lambda_j} \right)^{\alpha+\beta}+\left(\frac{\lambda_i}{\lambda_j} \right)^{1-\alpha-\beta} \right\}^2 - \left\{ \left(\frac{\lambda_i}{\lambda_j} \right)^{\alpha}+\left(\frac{\lambda_i}{\lambda_j} \right)^{1-\alpha}+\left(\frac{\lambda_i}{\lambda_j} \right)^{\beta}+\left(\frac{\lambda_i}{\lambda_j} \right)^{1-\beta} \right\}^2
$$
$$
\geq 16 \alpha \beta \left( \frac{\lambda_i}{\lambda_j}-1 \right)^2.
$$
Then we have 
\begin{eqnarray}
&   & \{ \lambda_i+\lambda_j+f_{\alpha+\beta}(\lambda_i,\lambda_j)-f_{\alpha}(\lambda_i,\lambda_j)-f_{\beta}(\lambda_i,\lambda_j) \} \nonumber \\
&   & \times \{ \lambda_i+\lambda_j+f_{\alpha+\beta}(\lambda_i,\lambda_j)+f_{\alpha}(\lambda_i,\lambda_j)+f_{\beta}(\lambda_i,\lambda_j) \}\nonumber \\
& = & (\lambda_i+\lambda_j+f_{\alpha+\beta}(\lambda_i,\lambda_j))^2-(f_{\alpha}(\lambda_i,\lambda_j)+f_{\beta}(\lambda_i,\lambda_j))^2 \nonumber \\
& \geq & 16 \alpha \beta (\lambda_i-\lambda_j)^2. \label{eq:num3-9}
\end{eqnarray}
Since 
\begin{eqnarray*}
Tr[\rho[A,B]] & = & Tr[\rho[A_0,B_0]] \\
& = & 2i Im Tr[\rho A_0B_0] \\ 
& = & 2i Im \sum_{i<j} (\lambda_i-\lambda_j) \langle \phi_i|A_0|\phi_j \rangle \langle \phi_j|B_0|\phi_i \rangle \\
& = & 2i \sum_{i<j} (\lambda_i-\lambda_j) Im \langle \phi_i|A_0|\phi_j \rangle \langle \phi_j|B_0|\phi_i \rangle, 
\end{eqnarray*}
\begin{eqnarray*}
|Tr[\rho[A,B]]| & = & 2|\sum_{i<j} (\lambda_i-\lambda_j) Im \langle \phi_i|A_0|\phi_j \rangle \langle \phi_j|B_0|\phi_i \rangle | \\
& \leq & 2 \sum_{i<j} |\lambda_i-\lambda_j||Im \langle \phi_i|A_0|\phi_j \rangle \langle \phi_j|B_0|\phi_i \rangle |. 
\end{eqnarray*}
Then we have 
$$
|Tr[\rho[A,B]]|^2 \leq 4 \left\{ \sum_{i<j} |\lambda_i-\lambda_j|| Im \langle \phi_i|A_0|\phi_j \rangle \langle \phi_j|_0|\phi_i \rangle | \right\}^2.
$$
By (\ref{eq:num3-9}) and Schwarz inequality, 
\begin{eqnarray*}
&   & \alpha \beta |Tr[\rho[A,B]]|^2 \\
& \leq & 4 \alpha \beta \left\{ \sum_{i<j} |\lambda_i-\lambda_j|| Im \langle \phi_i|A_0|\phi_j \rangle \langle \phi_j|B_0|\phi_i \rangle | \right\}^2 \\
& = & \frac{1}{4} \left\{ \sum_{i<j} 4 \sqrt{\alpha \beta}|\lambda_i-\lambda_j|| Im \langle \phi_i|A_0|\phi_j \rangle \langle \phi_j|B_0|\phi_i \rangle | \right\}^2 \\
& \leq & \frac{1}{4} \left\{ \sum_{i<j} 4 \sqrt{\alpha \beta}|\lambda_i-\lambda_j| | \langle \phi_i|A_0|\phi_j \rangle || \langle \phi_j|B_0|\phi_i \rangle | \right\}^2 \\
& \leq & \frac{1}{4} \left\{ \sum_{i<j} \{ (\lambda_i+\lambda_j+f_{\alpha+\beta}(\lambda_i,\lambda_j))^2-(f_{\alpha}(\lambda_i,\lambda_j)+f_{\beta}(\lambda_i,\lambda_j)^2 \}^{1/2} | \langle \phi_i|A_0|\phi_j \rangle || \langle \phi_j|B_0|\phi_i \rangle | \right\}^2 \\
& \leq & \frac{1}{2} \sum_{i<j} \{ \lambda_i+\lambda_j+f_{\alpha+\beta}(\lambda_i,\lambda_j)-f_{\alpha}(\lambda_i,\lambda_j)-f_{\beta}(\lambda_i,\lambda_j) \} | \langle \phi_i|A_0|\phi_j \rangle |^2 \\
&   &  \times \frac{1}{2} \sum_{i<j} \{ \lambda_i+\lambda_j+f_{\alpha+\beta}(\lambda_i,\lambda_j)+f_{\alpha}(\lambda_i,\lambda_j)+f_{\beta}(\lambda_i,\lambda_j) \} | \langle \phi_i|B_0|\phi_j \rangle |^2.
\end{eqnarray*}
Then we have 
$$
I_{\rho,\alpha,\beta}(A) J_{\rho,\alpha,\beta}(B) \geq \alpha \beta |Tr[\rho [A,B]]|^2.
$$
We also have 
$$
I_{\rho,\alpha,\beta}(B) J_{\rho,\alpha,\beta}(A) \geq \alpha \beta |Tr[\rho [A,B]]|^2.
$$
Hence we have the final result (\ref{eq:num3-1}). \ \hfill $\Box$ 

\begin{remark}
We remark that (\ref{eq:num2-10}) is derived by putting $\beta = 1-\alpha$ in (\ref{eq:num3-1}). Then Theorem \ref{th:theorem3-1} 
is a generalization of Theorem \ref{th:theorem2-1} given in \cite{Ya:WYD}. 
\label{re:remark3-1}
\end{remark}


\begin{thebibliography}{99}

\bibitem{Bo:some} J.C.Bourin, 
{\it Some inequalities for norms on matrices and operators}, Linear Algebra and its Applications, 
vol.292(1999), pp.139-154. 
\bibitem{CaLu:convex} L.Cai and S.Luo, 
{\it On convexity of generalized Wigner-Yanase-Dyson information}, 
Lett. Math. Phys., vol.83(2008), pp.253-264.
\bibitem{Fu:trace} J.I.Fujii, 
{\it A trace inequality arising from quantum information theory}, Linear Algebra and its 
Applications, vol.400(2005), pp.141-146.
\bibitem{He:unq} W.Heisenberg,
{\it \"Uber den anschaulichen Inhat der quantummechanischen Kinematik und Mechanik}, 
Zeitschrift f\"ur Physik, vol.43(1927), pp.172-198.
\bibitem{Ko:matrix} H.Kosaki, 
{\it Matrix trace inequality related to uncertainty principle}, Internatonal 
Journal of Mathematics, vol.16(2005), pp.629-646.
\bibitem{Li:convex} E.H.Lieb, 
{\it Convex trace functions and the Wigner-Yanase-Dyson conjecture}, 
Adv. Math., vol.11(1973), pp.267-288.
\bibitem{Lu:Hei} S.Luo, 
{\it Heisenberg uncertainty relation for mixed states}, Phys. Rev. A, vol.72(2005), 
p.042110.
\bibitem{LuZh:skew} S.Luo and Q.Zhang, 
{\it On skew information}, IEEE Trans. Information Theory, vol.50(2004), 
pp.1778-1782, and {\it Correction to "On skew information"}, IEEE Trans. 
Information Theory, vol.51(2005), p.4432.
\bibitem{WY:inf} E.P.Wigner and M.M.Yanase, 
{\it Information content of distribution}, Proc. Nat. Acad. Sci. U,S,A., 
vol.49(1963), pp.910-918.
\bibitem{YaFuKu:gen} K.Yanagi, S.Furuichi and K.Kuriyama, 
{\it A generalized skew information and uncertainty relation}, IEEE Trans. 
Information Theory, vol.51(2005), pp.4401-4404.
\bibitem{FuYaKu:trace} S.Furuichi, K.Yanagi and K.Kuriyama, 
{\it Trace inequalities on a generalized Wigner-Yanase skew information}, 
J. Math. Anal. Appl., vol.356(2009), pp.179-185.
\bibitem{Ya:WYD} K.Yanagi, 
{\it Uncertainty relation on Wigner-Yanase-Dyson skew information}, 
J. Math. Anal. Appl., vol.365(2010), pp.12-18.

\end{thebibliography}
\end{document}